\documentclass[iop]{emulateapj}
\usepackage{amsmath,amssymb,multirow,color}

\usepackage{bm}
\usepackage{epsfig}
\usepackage{graphicx}
\usepackage{color}
\usepackage{subfigure}

\newcommand{\eq}[1]{Eq.~(\ref{#1})}
\newcommand{\eqs}[2]{Eqs.~(\ref{#1}--\ref{#2})}

\newcommand{\be}{\begin{equation}}
\newcommand{\ee}{\end{equation}}
\newcommand{\beq}{\begin{equation}}
\newcommand{\eeq}{\end{equation}}

\newcommand\bea{\begin{eqnarray}}
\newcommand\eea{\end{eqnarray}}

\renewcommand{\(}{\left(}
\renewcommand{\)}{\right)}

\begin{document}
\title{Collisionless Reconnection in Magnetohydrodynamic and Kinetic Turbulence}
\author{ Nuno\ F.\ Loureiro$^{1}$, Stanislav Boldyrev$^{2,3}$}
\affil{${~}^1$Plasma Science and Fusion Center, Massachusetts Institute of Technology, Cambridge MA 02139, USA}
\affil{${~}^2$Department of Physics, University of Wisconsin, Madison, WI 53706, USA\\
${~}^3$Space Science Institute, Boulder, Colorado 80301, USA}

\date{\today}

\begin{abstract}
It has recently been proposed \cite[][]{loureiro2017,mallet_disruption_2017} that the inertial interval in magnetohydrodynamic (MHD) turbulence is terminated at small scales not by a Kolmogorov-like dissipation region, but rather by a new sub-inertial interval mediated by tearing instability. 
However, many astrophysical plasmas are nearly collisionless so that the MHD approximation is not applicable to turbulence at small scales. 
In this Letter, we propose the extension of the theory of reconnection-mediated turbulence to plasmas which are so weakly collisional that the reconnection occurring in the turbulent eddies is caused by electron inertia rather than by resistivity. 
We find that the transition scale to reconnection-mediated turbulence depends on the plasma beta and on the assumptions of the plasma turbulence model. 
However, in all cases analyzed, the energy spectra  in the reconnection-mediated interval range from $E(k)dk\propto k^{-8/3}dk$ to $E(k)dk\propto k^{-3}dk$. 
\end{abstract}
\maketitle


\section{Introduction}
\label{sec:introduction}
In many astrophysical flows, such as those governing stellar coronae and winds, the dynamics of planetary magnetospheres, the structures in the warm interstellar medium, and many others, the dissipation scales are so small that the flows become turbulent over a broad range of scales. 
At scales much larger than the plasma microscales (the particles' gyroradii and skin depths), the dynamics can be reasonably well approximated by single-fluid magnetohydrodynamics (MHD) \cite[e.g.,][]{biskamp2003,davidson2017}.
Some fundamental processes of plasma energetics, such as plasma heating, particle acceleration, magnetic field reconnection, however, depend on plasma turbulence at the kinetic scales \cite[e.g.,][]{kulsrud_plasma_2005,chen2016}. 
At such scales, the MHD description is not adequate, and the need arises of extending the relatively well-developed theory of MHD turbulence to the small, kinetic scales.

Recently, it has been suggested that magnetic reconnection may be a critical element of MHD turbulence~\citep{loureiro2017,mallet_disruption_2017,boldyrev_2017}.
This realization arises from the observation, motivated by the dynamic alignment picture~\cite[][]{boldyrev_spectrum_2006}, that turbulent eddies become progressively more unstable to the tearing mode instability (which drives reconnection) as their characteristic scale, $\lambda$, gets smaller. 
 Those authors have thus proposed that the role of reconnection in MHD turbulence can be quantified by comparing the characteristic timescales of the two processes.
A typical turbulent eddy at scale $\lambda$ lasts an amount of time (the eddy turnover time) that is itself a function of $\lambda$. Similarly, a reconnection event at scale $\lambda$ occurs on a timescale which depends in some nontrivial way on $\lambda$. 
If the time associated with reconnection decreases with the scale of the eddy faster than the eddy turnover time does, then, given a large enough inertial interval,  reconnection inevitably becomes important below a certain critical scale $\lambda_c$. 
This scale is thus defined as the scale below which reconnection becomes faster than the turbulence --- for this is the only way to ensure that reconnection has time to occur before the eddies are destroyed by the turbulence. 

In \cite[][]{loureiro2017} it was proposed that the scale-invariant energy cascade should persist in the reconnection-mediated interval, and that the energy should have a power-law Fourier spectrum. 
This concept was further developed in \cite[][]{boldyrev_2017} and along somewhat different lines of reasoning in \cite[][]{mallet_disruption_2017}.\footnote{Interestingly, \citet{huang_turbulent_2016} observed a similar power-law spectrum in turbulence driven by plasmoid instability. Although in their setting the reconnecting magnetic profiles were not generated by turbulence, their numerical observations may have presented the first glimpse at the phenomenon studied in these works.}
However, these interesting new developments cannot be directly applied to very weakly-collisional turbulent astrophysical environments such as those mentioned earlier. There are two reasons for that. 
{First, even if the eddy scale itself is larger than the kinetic scales, the tearing instability in such an eddy leads to the formation of finer scales (a boundary layer) which are, almost inevitably, in the kinetic regime. 
Indeed, for realistic values of the resistivity, one finds that the mechanism breaking the frozen flux condition, and thus enabling reconnection, is likely to be the electron inertia, rather than the resistivity. 
Second, turbulence in very weakly-collisionless plasmas extends to sub-ion scales; it is possible, as we will show, that reconnection may only become important at those scales.

This Letter presents the first attempt at extending these recent ideas on the role of reconnection in turbulence to accommodate kinetic physics.

\section{The tearing mode in a strongly magnetized, collisionless plasma}
\label{sec:tearing}
Let us begin by analyzing the case of a low-beta plasma, $m_e/m_i\ll \beta\ll 1$, where $\beta = 8\pi n T /B_0^2$ (here we  implicitly assume the ion and electron temperatures to be comparable, though not necessarily equal) and $B_0$ is a large-scale uniform magnetic field whose presence we assume. 
Small magnetic fluctuations in the direction normal to $B_0$ will be denoted as $b$. 
We will {first} consider plasma fluctuations with typical scales $a$ such that
\begin{eqnarray}
\label{scale_a}
a\gg\rho_i\sim\rho_s,
\end{eqnarray}
where $\rho_i$ is the ion gyroscale and $\rho_s$ is the ion acoustic scale.\footnote{In studies of turbulence the eddy scale $a$ is typically denoted by $\lambda$. We however keep in this section the notation traditional for the reconnection literature.} 
For simplicity, and without loss of generality, we assume an hydrogen plasma, which is one of the most relevant cases in astrophysical applications. As we discuss in more detail in the following section, it is reasonable to assume that the magnetic profile of a turbulent eddy at MHD scales resembles a current sheet with thickness~$a$~\cite[e.g.,][]{boldyrev_spectrum_2006,mallet_statistical_2016}.

A turbulent eddy with such a magnetic profile will reconnect if it is unstable to the tearing mode~\citep{FKR}. In this case, the magnetic profile tends to develop a singularity characterized by an inverse scale $\Delta^\prime$, which is a fundamental parameter of the tearing instability.  The other crucial parameter, the size of inner boundary layer of the tearing mode, $\delta_{in}$, is much smaller than the scale of the reconnecting magnetic profile. Due to Eq.~(\ref{scale_a}), the eddy scale $a$ belongs to the MHD regime of plasma turbulence. We, however, assume that the inner scale $\delta_{in}$ belongs to the kinetic range,
\begin{eqnarray}
\label{delta_in}
\delta_{in}\ll \rho_i\sim\rho_s.
\end{eqnarray}
It is inequality (\ref{delta_in}) that ensures that kinetic effects (electron inertia in this case) become important at the inner scale of the tearing mode. 
This inequality distinguishes our kinetic theory from the MHD theory of reconnection-mediated turbulence developed in \cite[][]{loureiro2017,boldyrev_2017}.  

Under these conditions, the linear growth rate of the tearing instability of such a magnetic field configuration scales as follows (see~\cite[][]{zocco_reduced_2011} and references therein). 
For low $\Delta'$ (i.e., for $\Delta'\delta_{in}\ll 1$) we have:
\be
  \label{eq:low-dprime}
  \gamma \sim k v_A d_e \rho_s \Delta' / a,
\ee
whereas for large $\Delta'$ (i.e., $\Delta'\delta_{in}\gtrsim 1$)
\be
  \label{eq:large-dprime}
  \gamma\sim k v_A d_e^{1/3} \rho_s^{2/3}/a.
\ee
In these expressions, $v_A$ denotes the Alfv\'en speed based on the reconnecting magnetic field $b_{a}$, $k$ is the wavenumber of the perturbation parallel to the reconnecting field, $d_e\equiv c/\omega_{pe}$ is the electron skin depth (with $\omega_{pe}$ the electron plasma frequency).\footnote{These expressions are valid for finite ion temperature; this, however, only affects the numerical prefactors that multiply these expressions~\citep{zocco_reduced_2011}, and so does not affect the order of magnitude derivations that follow.}

The instability parameter $\Delta'$ is obtained by solving the outer region (MHD) equation. It is a function of the wavenumber $k$, but its specific scaling with $k$ (for $ka\ll 1$) depends on the functional form of the reconnecting magnetic field.
For the usual Harris magnetic field profile~\citep{harris_1962} it is $\Delta'a\sim 1/(ka)$. In the Harris profile, the field reverses direction on the scale $a$, but its plateau region is much longer than $a$ (strictly speaking, it is infinite). In the other limiting case, where the scales of the field reversal and the field plateau are comparable (say a sinusoidal profile), the scaling is $\Delta'a\sim 1/(ka)^2$.
Because both scalings are, {\it a priori}, possible in a turbulent eddy, we will keep this dependence generic by assuming simply that $\Delta'a\sim 1/(ka)^n$, where $1<n\leq 2$ is a parameter (not necessarily integer).

The usual procedure to find the wavenumber of the fastest growing tearing mode is to equate the expressions for the growth rate and solve for $k_{max}$. This gives
\begin{eqnarray}
k_{max}^{(n)}\sim d_e^{2/(3n)}\rho_s^{1/(3n)}a^{-2-1/n},
\label{k_n}
\end{eqnarray}
which corresponds to the maximum growth rate
\begin{eqnarray}
\label{eq:gmax_low_beta}
\gamma_{max}^{(n)}\sim v_A d_e^{(1+2/n)/3}\rho_s^{(2+1/n)/3}a^{-2-1/n}.
\label{gamma_n}
\end{eqnarray}
For the Harris sheet-like configuration, the $n\to 1$ case, the fastest growing mode and the corresponding maximal growth rate are:
\bea
k_{max}^{(1)} &\sim & d_e^{2/3}\rho_s^{1/3}/a^2,\\
\gamma_{max}^{(1)}&\sim & v_A d_e \rho_s/a^3. 
\end{eqnarray}
In the limiting case $n=2$, the fastest growing mode and the associated growth rate are:
\bea
k_{max}^{(2)}&\sim & d_e^{1/3}\rho_s^{1/6} a^{-3/2},\\
\gamma_{max}^{(2)}&\sim & v_A d_e^{2/3}\rho_s^{5/6}a^{-5/2}.  
\eea

\section{Kinetic Reconnection in MHD turbulence}
\label{sec:rec_in_turb}
Let us now apply these scalings in the context of MHD turbulence as described by \cite{boldyrev_spectrum_2006}.
Therefore, we envision a tearing-unstable magnetic profile whose characteristic scale, identified as $a$ above, is the width of the turbulent eddy, $\lambda$. Correspondingly, the length of the current sheet, $L_{cs}$, is the other field-perpendicular eddy dimension, 
\be
\label{eq:xi_lambda}
\xi\sim L (\lambda/L)^{3/4}, 
\ee
where $L$ is the outer scale of the turbulence.
Likewise, the Alfv\'en velocity that appears in the scalings above is identified with the Alfv\'en velocity at scale $\lambda$, i.e., 
\be
\label{eq:va_lambda}
v_A\rightarrow v_{A,\lambda}\sim v_{A,0}(\lambda/L)^{1/4},
\ee
with  $v_{A,0}$ is the Alfv\'en velocity at the outer scale $L$.
The eddy turnover time at scale $\lambda$ is 
\be
\label{eq:tau_lambda}
\tau\sim \lambda^{1/2} L^{1/2}/v_{A,0}.
\ee
The following calculations hold provided that $\lambda$ remains larger than any kinetic scales.

For the general magnetic profile, the transition scale between the inertial and reconnection-dominated intervals is given by the criterion
\be
\label{eq:rec_onset}
\gamma_{max}^{(n)}\tau\sim 1.
\ee
This yields
\begin{eqnarray}
\lambda_{cr}^{(n)}/L\sim (d_e/L)^{\frac{1}{3}\(1+\frac{2}{n}\)/\(\frac{5}{4}+\frac{1}{n}\)}(\rho_s/L)^{\frac{1}{3}\(2+\frac{1}{n}\)/\(\frac{5}{4}+\frac{1}{n}\)}.\quad
\label{lambda_n}
\end{eqnarray}
In the limiting case $n=1$, this relation becomes\footnote{A. Mallet, A. Schekochihin, and B. Chandran have informed us in a private communication that they have independently arrived at the same expression for the transition scale.}
\begin{eqnarray}
\lambda_{cr}^{(1)}/L \sim (d_e/L)^{4/9}(\rho_s/L)^{4/9}.
\end{eqnarray}
This expression is valid provided that $\lambda_{cr} \gg \rho_s$, i.e., 
\begin{eqnarray}
\rho_s/L\ll (m_e/m_i)^2\beta_e^{-2}.
\label{restriction1}
\end{eqnarray}
where we have used the definition $d_e \sim \rho_s (m_e/m_i)^{1/2}\beta_e^{-1/2}$.

For an $n=2$ type magnetic profile, we instead obtain
\be
\label{eq:lcr_n2}
\lambda_{cr}^{(2)}/L \sim (d_e/L)^{8/21}(\rho_s/L)^{10/21}.
\ee
The validity condition $\lambda_{cr} \gg \rho_s$ now implies
\be
{\rho_s}/{L} \ll ({m_e}/{m_i})^{4/3}\beta_e^{-4/3}.
\label{restriction2}
\ee
Both limits of our model provide similar transition scales and rather weak parameter restrictions for the reconnection-mediated turbulence in a low-beta plasma. For instance, for the plasma in the solar corona, where one expects $\beta_e\sim 0.01$ at the distance of $\sim 10$ solar radii from the sun \cite[e.g.,][]{chandran2011}, we obtain from Eq.~(\ref{restriction1}):
\be
{\rho_s}/{L} \ll 3\times 10^{-3},
\ee
while the restriction corresponding to the other limiting case (\ref{restriction2}) yields
\be
{\rho_s}/{L} \ll 10^{-1}.
\ee
In both cases a Hydrogen plasma is assumed. We may therefore expect that the inertial interval of the coronal Alfv\'enic turbulence should transform into the reconnection-mediated interval at small scales.

\section{Turbulence Spectrum}
To obtain the energy spectrum in the reconnection-mediated range, we proceed as in ~\cite{boldyrev_2017}. 
Specifically, we assume that a consequence of the tearing mode becoming nonlinear is that the eddy evolution rate ($\gamma_{nl}$) then becomes enslaved to that of the mode, i.e., 
\be
\label{eq:eddy_rec_gamma}
\gamma_{nl}\sim\gamma_t.
\ee
{As in the MHD case, it is known from theoretical and numerical studies that the growth rate of the kinetic tearing mode discussed in Section~\ref{sec:tearing} remains unchanged from its linear value as the mode enters the nonlinear regime~\cite[e.g.,][]{wang_nonlinear_1993,porcelli_recent_2002}} .

Thus, let us define the energy cascade rate $\epsilon=V_{A0}^3/L_0$ and assume it to be independent of scale ($\lambda$) both in the inertial and reconnection ranges. Dimensionally, we have
\be
\gamma_{nl} \sim \epsilon/v_{A\lambda}^2.
\ee
Then, imposing $\gamma_{nl}\sim \gamma_t$, we obtain, for $n=1$ type magnetic profiles,
\be
v_{A\lambda}\sim\epsilon^{1/3}\lambda d_e^{-1/3} \rho_s^{-1/3},
\ee
from which one easily finds
\be
\label{eq:spectrum_lowbeta_n1}
E(k_\perp) dk_\perp \sim \epsilon^{2/3} d_e^{-2/3} \rho_s^{-2/3} k_\perp^{-3}dk_\perp.
\ee

Similarly, it is easy to see that for a type $n=2$ magnetic profile one has
\be
v_{A\lambda}\sim \epsilon^{1/3}\lambda^{5/6}d_e^{-2/9}\rho_s^{-5/18},
\ee
corresponding to the energy spectrum
\be
\label{eq:spectrum_lowbeta_n2}
E(k_\perp) dk_\perp \sim \epsilon^{2/3} d_e^{-4/9} \rho_s^{-5/9} k_\perp^{-8/3}dk_\perp.
\ee

According to our results, the energy spectrum of Alfv\'enic turbulence mediated by kinetic reconnection should therefore range from $k_{\perp}^{-8/3}$ to $k_{\perp}^{-3}$.

\section{Ultralow beta limit}
The limit when plasma beta is so low that $\beta_e\ll m_e/m_i$ can similarly be considered; this is relevant for, e.g., the Earth's magnetosphere \cite[e.g.,][]{chaston2008}. 

The calculation proceeds as in the previous sections. From~\cite{zocco_reduced_2011}, we find that in this limit the fastest growing tearing mode wavenumber and corresponding growth rate are
\begin{align}
k_{max}^{(n)} &\sim d_e^{1/n}a^{-1-1/n},\\
\gamma_{max}^{(n)} &\sim d_e^{1+1/n} v_A a^{-2-1/n}.
\end{align}

Application of the criterion stated by \eq{eq:rec_onset} yields the following critical scale for reconnection onset:
\be
\lambda_{cr}^{(n)}/L \sim (d_e/L)^{(1+\frac{1}{n})/(\frac{5}{4} + \frac{1}{n})}.
\ee
The two limiting cases of interest of this expression are $n=1$, for which we find
\be
\lambda_{cr}^{(1)}/L \sim (d_e/L)^{8/9};
\label{lambda_lb_1}
\ee
and $n=2$, which yields
\be
\label{lambda_lb_2}
\lambda_{cr}^{(2)}/L \sim (d_e/L)^{6/7}.
\ee

The validity of this analysis requires $\lambda_{cr}^{(n)}>d_e$, which in both cases reduces to $d_e<L$, a condition that is trivially satisfied.

As in the previous section, we can compute the energy spectra in this regime assuming that the growth rate of the tearing mode in the early nonlinear stage remains unchanged from its linear value. We obtain:
\begin{align}
E^{(1)}(k_\perp)& dk_\perp \sim \epsilon^{2/3} d_e^{4/3} k_\perp^{-3} dk_\perp,\\
E^{(2)}(k_\perp)& dk_\perp \sim \epsilon^{2/3} d_e^{-1} k_\perp^{-8/3} dk_\perp.
\end{align}
We see that the spectral slopes are unaltered from the previous expressions.
We also note that the transition scales (\ref{lambda_lb_1}) and (\ref{lambda_lb_2}) are always larger than the electron inertial scale $d_e$. This may be consistent with the slightly larger than $d_e$ scale of the Alfv\'enic spectral break observed in the Earth's magnetosphere turbulence \cite[][Fig. 3]{chaston2008}.

\section{Large $\beta$}
\label{sec:large_beta}
Another case of interest are plasmas where $\beta_e\sim1$. This can be considered using the approximate two-fluid tearing mode scalings derived in \citep{fitzpatrick2004,fitzpatrick2007}.\footnote{The equations in \citep{fitzpatrick2004,fitzpatrick2007} are formally derived in the cold-ion limit. However, usually the inclusion of finite ion temperature does not modify the scalings, only numerical prefactors; so, it is possible that the results in this section apply equally to cases where $\tau\sim1$.} Again, the procedure is entirely similar, so we simply state the key results.

The tearing mode dispersion relations in this regime are:
\be
\gamma \sim k v_A (d_i/a) \Delta' d_e,
\ee
for low $\Delta'$; and 
\be
\gamma \sim k v_A (d_i/a)^{3/5}(d_e/a)^{2/5},
\ee
at large $\Delta'$.
The most unstable mode and corresponding growth rate is:
\bea
k_{max}^{(n)} &\sim& d_e^{3/(5n)} d_i^{2/(5n)} a^{-1-1/n},\\
\label{eq:gmax_high_beta}
\gamma_{max}^{(n)} &\sim& v_A  d_e^{(3+2n)/(5n)}  d_i^{(2+3n)/(5n)} a^{-2-1/n}.
\eea
Note that \eq{eq:gmax_high_beta} exhibits the same dependence on $a$ as \eq{eq:gmax_low_beta}.

Using these relationships, we find that the critical eddy size for transition to the reconnection-mediated turbulence range is
\be
\lambda_{cr}^{(n)}/L \sim (d_e/L)^{4/5 (3+2n)/(4+5n)} (d_i/L)^{4/5 (2+3n)/(4+5n)},\quad
\ee
which has the following two limiting cases:
\be
\lambda_{cr}^{(1)}/L \sim (d_e/L)^{4/9} (d_i/L)^{4/9},
\ee
valid if
\be
\label{r_hb_1}
d_i/L \ll (m_e/m_i)^2;
\ee
and
\be
\lambda_{cr}^{(2)}/L \sim (d_e/L)^{2/5} (d_i/L)^{16/35},
\ee
valid if
\be
\label{r_hb_2}
d_i/L \ll (m_e/m_i)^{7/3}.
\ee
The corresponding spectra can be easily obtained:
\bea
E^{(1)}(k_\perp)dk_\perp \sim \epsilon^{2/3}d_e^{-2/3} d_i^{-2/3} k_\perp^{-3} dk_\perp,\\
E^{(2)}(k_\perp) dk_\perp \sim \epsilon^{2/3}d_e^{-7/15} d_i^{-8/15} k_\perp^{-8/3} dk_\perp.
\eea
These predictions for the energy spectra exhibit the same $k_\perp$ power law indices as those obtained earlier at low $\beta$, Eqs.~(\ref{eq:spectrum_lowbeta_n1}, \ref{eq:spectrum_lowbeta_n2}) --- the reason being that the growth rate of the tearing mode in this regime. \eq{eq:gmax_high_beta}, has the same dependence on the magnetic shear length $a$ (equivalently, $\lambda$) as \eq{eq:gmax_low_beta}. We however note that conditions (\ref{r_hb_1}) and (\ref{r_hb_2}) are very stringent and may, in fact, imply that in plasmas where $\beta_e\sim 1$, the transition to the reconnection range cannot happen in the MHD-scale interval.} 

\section{Reconnection in the kinetic turbulence range}
\label{sec:kinetic}	
One would now like to extend these ideas to the kinetic turbulence range, when the eddies are on sub-ion scales, i.e.,  $\lambda < \max(\rho_i,\rho_s)$.
This is, however, nontrivial, because our understanding of kinetic-scale turbulence is much less developed than that of MHD turbulence.
In particular, we are not familiar with an analytical theory that offers the kinetic equivalent of \eqs{eq:xi_lambda}{eq:tau_lambda}, implying, therefore, that we cannot know whether the tendency to 
develop current sheets that is present in the MHD range remains true in the kinetic range. 
However, numerical simulations and observations~\cite[e.g.,][]{boldyrev12b,wan_etal2012,tenbarge_current_2013,chen_etal2015,wan_etal2016,cerri_reconnection_2017} 
do show evidence for current sheet formation at such scales, and so perhaps it is legitimate to assume that current sheets remain the fundamental units of sub-ion scales turbulence.

Let us then assume that this is indeed so. 
There are two options: either the critical scale for onset of the reconnection range has been met at the MHD scales, i.e., $\lambda_c>\max(\rho_i,\rho_s)$, or it has not. 
In the latter case, we are not able to estimate it, because there is currently no theory to describe the eddy structure at those scales. 
Therefore, it remains to be seen whether reconnection may become important at the sub-proton scales. 
If, however, this is the case, 
we may compute the energy spectrum, provided that \eq{eq:eddy_rec_gamma} holds and that the tearing mode growth rate in the early nonlinear regime remains unchanged from its linear value.

We will address this question in the framework of the equations derived in \citep{chen2017}, valid at scales below the ion Larmor radius and assuming $\beta_i\gg \beta_e$ and $\beta_e\ll 1$ (Eqs. (19-20) of that reference). 
Such a regime may be relevant for the solar corona \cite[e.g.,][]{chandran2011}, hot accretion flows \cite[e.g.,][]{quataert98}, collisionless shocks \cite[e.g.,][]{treumann09,ghavamian13,chen2017}, etc.
The tearing mode calculation proceeds in the usual way; it is not hard to see that the most unstable tearing mode is such that $\Delta' \delta \sim 1$ and $\delta\sim d_e$.
This allows us to find immediately that: 
\begin{align}
k_{max}^{(n)} &\sim d_e^{1/n} a^{-1-1/n},\\
\gamma_{max}^{(n)} & \sim d_i d_e^{1/n} a^{-2-1/n}, \text{ if } \beta_i\gtrsim 1. \\
\gamma_{max}^{(n)} & \sim \rho_i d_e^{1/n} a^{-2-1/n}, \text{ if } \beta_i\ll 1.
\end{align}
The expected energy spectra follow straightforwardly:
\begin{align}
 E^{(1)}(k_\perp)& dk_\perp \sim \epsilon^{2/3} (m_i/m_e)^{-1/3}d_e^{-4/3} k_\perp^{-3},\\
 E^{(2)}(k_\perp)& dk_\perp \sim \epsilon^{2/3} (m_i/m_e)^{-1/3}d_e^{-1} k_\perp^{-8/3},
 \end{align}
if $\beta_i\sim 1$. In the opposite case of $\beta_i\ll 1$ the above expressions for the spectrum appear multiplied by the prefactor $\beta_i^{-1/3}$.

Again, interestingly, we observe that the $k_\perp$ dependence of the spectra are the same as in all cases considered above.
\section{Discussion and Conclusion}
\label{Discussion}
We have proposed that in collisionless plasmas, the inertial interval of Alfv\'enic turbulence can cross over to a reconnection-mediated interval at scales larger than the relevant plasma micro-scales, such as the ion-acoustic scale or the electron skin depth. 
We predict that depending on the parameters of the model, the magnetic energy spectrum in the reconnection-mediated interval can vary from $E(k)\propto k^{-8/3}$ to $E(k)\propto k^{-3}$, both in low beta plasmas (e.g.,  as the solar corona, interplanetary coronal mass ejections, planetary magnetospheres, etc.~\cite[e.g.,][]{chen14b,chaston2008,bale2016}) and plasmas with $\beta\sim 1$. 

We have suggested that our theory may be extended 
to the subproton-scale turbulence $\lambda < \rho_i$. 
Indeed, the turbulent fluctuations at such scales resemble current sheets \cite[e.g.,][]{boldyrev12b,wan_etal2012,tenbarge_current_2013,chen_etal2015,wan_etal2016,cerri_reconnection_2017}, although those turbulent structures (and generally turbulence at kinetic scales) are relatively less well understood than their Alfv\'enic counterpart. 
If we may assume that, similarly to the Alfv\'enic case, the dynamics at subproton scales are governed by the fastest growing tearing modes, the reconnection-dominated turbulence should have the same scaling $E(k)\propto k^{-8/3}$ to $E(k)\propto k^{-3}$ that we have derived for the Alfv\'enic case. 

Interestingly, the predicted spectral scaling is very close to the spectrum of turbulence $\approx -2.8$ measured in the $\beta\sim 1$ solar wind plasma below the ion-cyclotron scale \cite[e.g.,][]{alexandrova09,kiyani09a,chen10b,chen12a,sahraoui13a}. 
We should caution that the close proximity of the reconnection-mediated spectra of turbulence to the spectra derived from qualitatively different turbulence models (related to cascades of the kinetic-Alfv\'en or, possibly, whistler waves \cite[][]{howes08a,schekochihin09,chen10a,boldyrev12b}), may not allow one to discern what physical mechanism is dominant based solely on measurements of the spectral exponents.  
Numerical simulations and observations, however, offer increasing evidence that current sheets are important dynamic players in turbulence at both MHD and kinetic scales.
It thus seems conceivable to suppose that their presence affects the spectral properties of turbulence; this Letter presents the first theoretical analysis of how they may do so.


\paragraph{Acknowledgments.}
We are grateful to Christopher Chen for useful comments.
NFL was supported by the NSF-DOE Partnership in Basic Plasma Science and Engineering, award no. DE-SC0016215, and by NSF CAREER award no. 1654168. SB is partly supported by the National Science Foundation under the grant NSF AGS-1261659 and by the Vilas Associates Award from the University of Wisconsin - Madison. 
%

\end{document}